\documentclass[aps,pre,reprint,twocolumn,superscriptaddress,longbibliography,showpacs,floatfix]{revtex4-2}

\usepackage{graphicx}
\usepackage{amsmath}
\usepackage{amssymb}
\usepackage{epsf}
\usepackage{color}
\usepackage{mathtools}
\usepackage{bm}
\usepackage[colorlinks=true, citecolor=blue, linkcolor=blue, urlcolor=blue]{hyperref}

\begin{document}

\title{Identifying the sources of noise synergy and redundancy in the gene expression of feed-forward loop motif}

\author{Mintu Nandi}
\altaffiliation{Present Address: Universal Biology Institute, The University of Tokyo, 7-3-1 Hongo, Bunkyo-ku, Tokyo 113-0033, Japan}
\affiliation{Department of Chemistry, Indian Institute of Engineering Science and Technology, Shibpur, Howrah 711103, India.}

\author{Sudip Chattopadhyay}
\affiliation{Department of Chemistry, Indian Institute of Engineering Science and Technology, Shibpur, Howrah 711103, India.}

\author{Suman K Banik}
\affiliation{Department of Chemical Sciences, Bose Institute, EN 80, Sector V, Bidhan Nagar, Kolkata 700091, India.}

\begin{abstract}
The propagation of noise through parallel regulatory pathways is a characteristic feature of feed-forward loops in genetic networks. Although the contributions of the direct and indirect regulatory pathways of feed-forward loops to output variability have been well characterized, the impact of their joint action arising from their shared input and output remains poorly understood. Here, we identify an additional component of noise that emerges specifically from this convergent nature of the pathways. Using inter-gene correlations, we reveal the regulatory basis of the cross-interaction noise and interpret it as synergy or redundancy in noise propagation, depending on whether the combined pathways amplify or suppress fluctuations. Synergy typically arises in coherent feed-forward loops, whereas redundancy is common in incoherent ones. This framework not only accounts for previously observed differences in noise behavior across coherent and incoherent structures but also provides a generalizable strategy to connect network structure with stochastic gene regulation. Furthermore, by relating these synergy and redundancy to dynamical properties such as sign-sensitive delay or response acceleration, the framework offers a statistical lens to interpret the functional roles in cellular decision-making.
\end{abstract}

\date{\today}

\maketitle

\section*{Introduction}

\label{sec1}

\begin{figure*}[!t]
\includegraphics[width=2.0\columnwidth,angle=0]{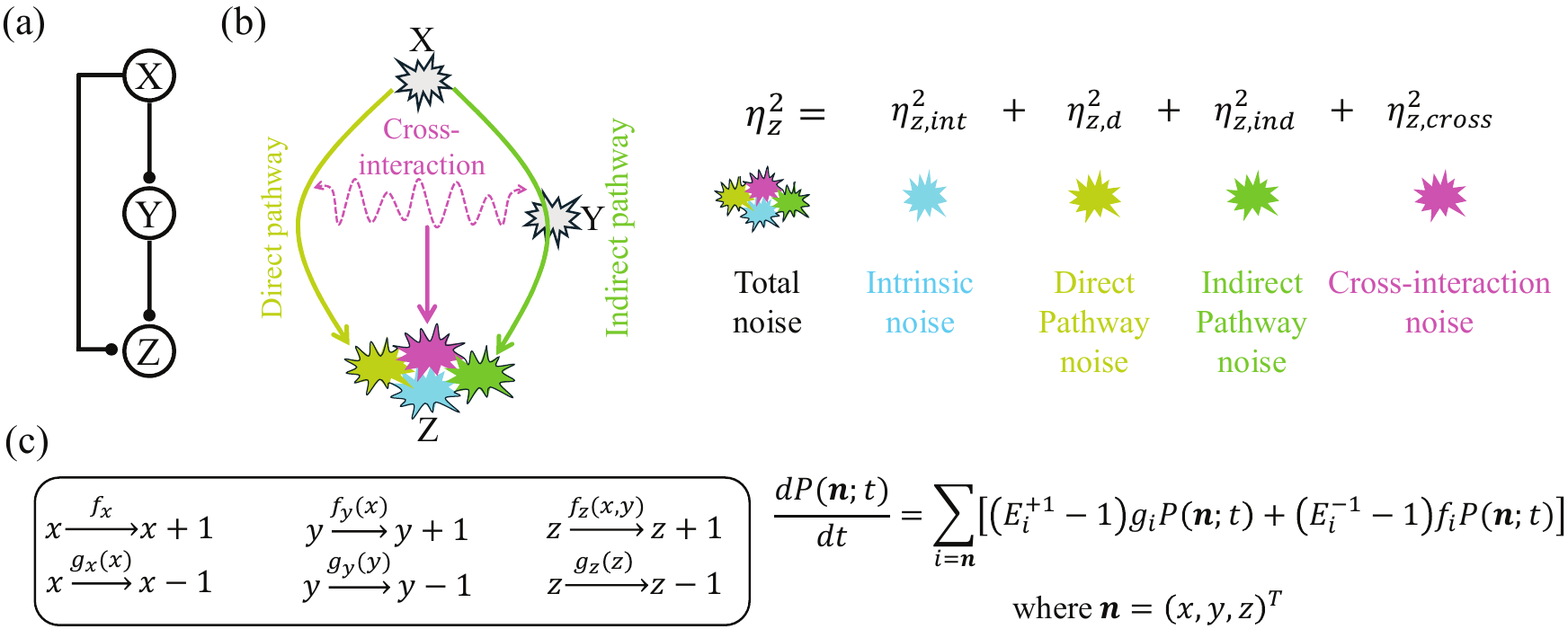}
\caption{\textbf{Schematics of dynamical and noise topology of FFL.} (a) A generic FFL containing three regulatory genes X, Y, and Z. (b) A schematic view of noise decomposition with identification of sources of each noise term. Notably, the cross-interaction noise is generated due to coupling between the two pathways in FFL. (c) Discrete dynamics of each gene of FFL with the corresponding master equation governing the temporal evolution of the joint probability $P({\bf n};t)$.
\label{f1}
}
\end{figure*}

Biological networks employ various regulatory strategies to control the propagation of noise \cite{Swain2004, Dublanche2006, Pedraza2008, Bleris2011}. A widely studied motif in such networks is the feed-forward loop (FFL) \cite{Alon2006}, composed of three genes connected through three regulatory edges. In FFLs, a source gene X regulates a target gene Z both directly and indirectly through an intermediate gene Y, forming two parallel signal propagation paths: X$\multimap$Z and X$\multimap$Y$\multimap$Z (Fig.~\ref{f1}a). These two signal propagation channels work together to make the motif functional. The pathways are integrated in the promoter of Z via either an AND or OR logic, which influences the overall response behavior \cite{Alon2007}.

FFLs are prevalent in the transcriptional networks of organisms such as \textit{Escherichia coli} and \textit{Saccharomyces cerevisiae} \cite{Shen-Orr2002, Mangan2003, Ma2004, Mangan2006}. Experimental studies have revealed their dynamic behavior, revealing their functional aspects \cite{Mangan2003, Kalir2004, Mangan2006, Kaplan2008, Bleris2011, Chen2013}. In parallel, theoretical work has elucidated their noise dynamics under different regulatory regimes \cite{Dunlop2008, Kittisopikul2010, Maity2015}. Some recent studies explored the robust nature of microRNA-mediated FFLs in mitigating external fluctuations by effectively filtering out propagated noise \cite{Li2009, Kittisopikul2010, Osella2011, Grigolon2016, Carignano2018}. A theory-experiment dialogue further investigated the noise-filtering behavior of synthetic CFFLs \cite{Pieters2021}. Beyond isolated FFLs, coupled FFLs have also received significant research attention \cite{Chalancon2012, Nandi2024_a}. A recent study compared the noise reduction capabilities of isolated and coupled FFLs, highlighting their relative performance under fluctuating conditions \cite{Chakravarty2021}.

Despite these efforts, most studies focus on dynamic or functional behavior. Less attention has been paid to how the FFL topology itself shapes noise propagation. In this paper, we focus on this structural aspect, aiming to understand how topological interactions govern the interplay between parallel regulatory paths and their contribution to noise regulation. 

In steady-state, the noise decomposition strategy provides the squared coefficient of variation (CV) of the output Z, $\eta_z^2 ~(=\sigma_z^2/\langle z\rangle^2)$ \cite{Nandi2021, Roy2023},
\begin{equation}
    \eta^2_z = \eta^2_{z,int} + \eta^2_{z,d} + \eta^2_{z,ind} + \eta^2_{z,cross},
    \label{eq3}
\end{equation}

\noindent where $\eta^2_{z, int}$ indicates the intrinsic contribution of noise from Z, while $\eta^2_{z, d}$ and $\eta^2_{z, ind}$ originate from noise propagation through direct and indirect pathways, respectively. Regardless of the structure of FFL, the cross-interaction noise, $\eta^2_{z, cross}$, arises from the integration of both direct and indirect pathways \cite{Nandi2021} (see Fig.~\ref{f1}b). This term has been understood as the combined effect of noise transmission from X to Z through these pathways \cite{Bruggeman2009}. Furthermore, it has been recognized as integrated noise that clarifies the amplification and attenuation characteristics of FFLs (see Fig.~S1) \cite{Nandi2021}. This additional component also plays an important role in highlighting the functional relevance of CFFL as an effective motif for information transmission \cite{Momin2020}. We note that this noise decomposition framework is general and applicable to all FFL architectures, regardless of whether the regulation at the Z promoter follows an AND or OR logic.

To date, the appearance of the cross-interaction noise has primarily been treated as a formal outcome of noise decomposition \cite{Bruggeman2009, Momin2020, Nandi2021, Roy2023}. A detailed characterization of this term is lacking in the literature. Accurate identification and characterization of the source(s) of the cross-interaction noise are essential. Although it is phenomenologically observed to originate from the integration of direct and indirect pathways, a deeper understanding of its origin is needed. In this article, our aim is to address these questions and provide a more comprehensive understanding of the role of this cross-interaction noise.

We approach these questions by first analyzing how the regulatory structure of the FFL induces correlations among the expression levels of X, Y, and Z. This helps us trace the origin of the cross-interaction noise in terms of specific inter-gene correlations. We then go beyond identifying the source by providing a functional interpretation of this noise term using the notion of synergy and redundancy. We demonstrate that this synergy–redundancy framework captures important steady-state features that reflect the dynamical roles of different FFL types. Our approach thus provides a new tool for linking network structure to gene expression variability and offers insights into how cells may tune their responses and maintain robustness under a changing environment.

\section*{Results}
\label{sec2}

We adopt the general noise decomposition introduced in Eq.~(\ref{eq3}) to quantify the contributions of different regulatory paths to the steady-state variability of the output gene Z. To explicitly evaluate each noise component, we derive the exact steady-state moments (variances and covariances) corresponding to the discrete stochastic dynamics shown in Fig.~\ref{f1}c (see \nameref{sec4}). This framework remains general and applies to all FFL architectures, regardless of whether the regulation at the Z promoter follows an AND or OR logic.

\subsection*{Characterization of distinct paths of gene-gene correlations}

We focus first on the steady-state covariances between gene products X, Y, and Z. By deriving analytical expressions for the normalized covariances, we aim to identify the distinct routes through which correlations propagate across the network. The normalized covariances, a measure of correlations, between gene pairs can be further decomposed into partial terms, each corresponding to a specific regulatory pathway. The explicit expressions for these normalized covariances are (see Eqs.~(S10–S12) of Supporting Information (SI)),
\begin{eqnarray}
    \eta_{xy}^2 &=& \underbrace{\frac{\Gamma_{xy}}{\langle y\rangle} \times f_{yx}^\prime}_{\rm X \multimap Y},
    \label{eq4} \\
    \eta_{xz}^2 &=& 
    \underbrace{\frac{\Gamma_{xz}}{\langle z\rangle} \times f_{zx}^\prime}_{\eta_{xz,d}^2:~ \rm X \multimap Z}
    +
    \underbrace{\frac{\Gamma_{xyz}^{(1)}}{\langle z\rangle} \times f_{yx}^\prime f_{zy}^\prime}_{\eta_{xz,ind}^2:~ \rm X \multimap Y \multimap Z},
    \label{eq5} \\
    \eta_{yz}^2 &=& 
    \underbrace{\frac{\Gamma_{yz}}{\langle z\rangle} \times f_{zy}^\prime}_{\eta_{yz,ind_1}^2:~ \rm Y \multimap Z} 
    +
    \underbrace{\frac{\Gamma_{xyz}^{(2)} \langle x\rangle}{\langle y\rangle \langle z\rangle} \times f_{yx}^{\prime^2} f_{zy}^\prime}_{\eta_{yz,ind_2}^2:~ \rm X \multimap Y \multimap Z}
    \nonumber \\
     && 
    + \underbrace{\frac{\Gamma_{xyz}^{(3)} \langle x\rangle}{\langle y\rangle \langle z\rangle} \times f_{yx}^\prime f_{zx}^\prime}_{\eta_{yz,cross}^2:~ \rm \{X \multimap Y\} \cup \{X \multimap Z\}},
    \label{eq6}
\end{eqnarray}

\noindent where $\Gamma$s are functions of the separation of time scales of gene products X, Y, and Z. In these expressions, $f_{ij}^\prime$ denotes the regulatory function of the $j \multimap i$ interaction. It is defined as the partial derivative of the synthesis function of the $i$th protein $f_i$ with respect to the population of $j$ in steady-state.  Its presence in the covariance terms helps identify the regulatory pathways responsible for inter-gene correlations in FFLs.

\begin{figure*}[!t]
\includegraphics[width=2.0\columnwidth,angle=0]{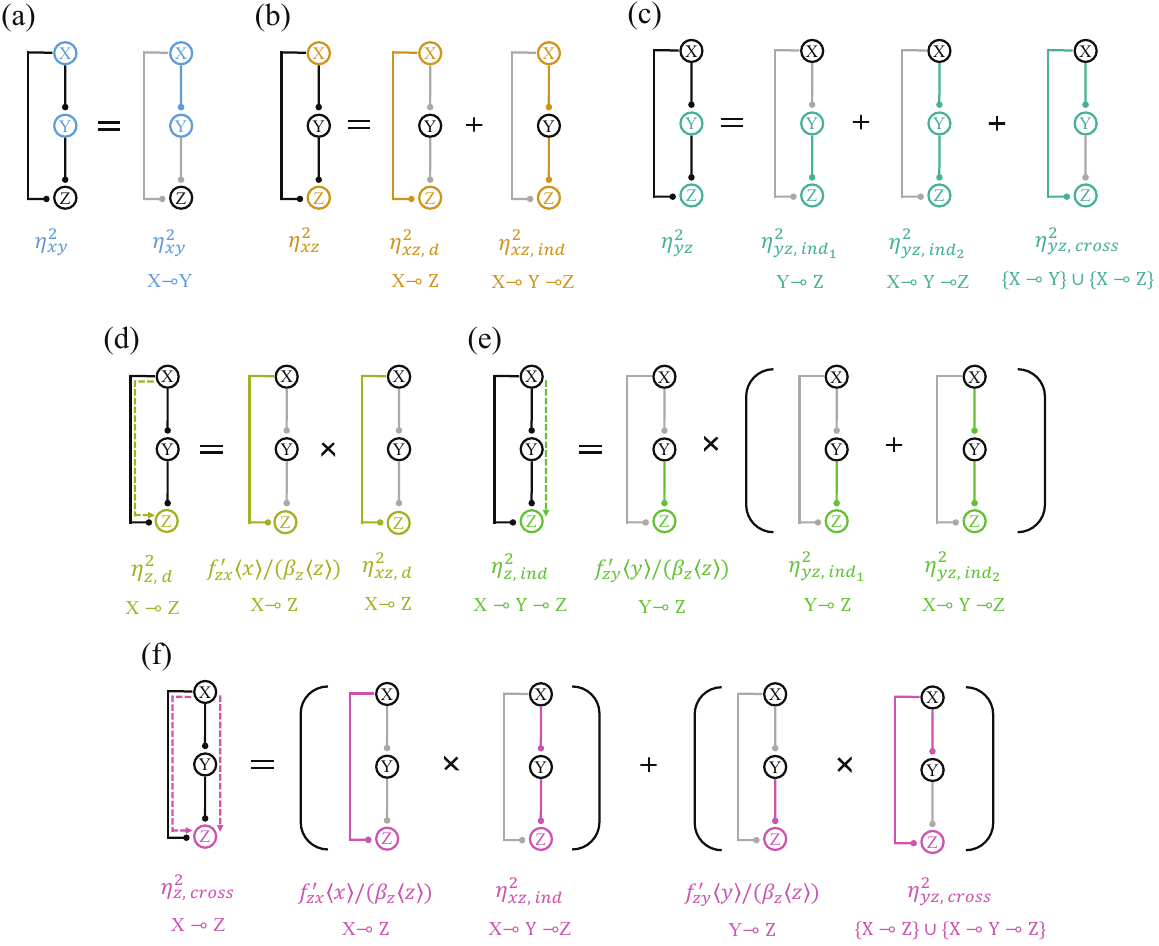}
\caption{\textbf{Graph-based illustration of gene–gene correlation decomposition and decomposed output noise components.}
\textit{Gene-gene correlation}: Colored nodes indicate correlated gene pairs, and the corresponding colored edges highlight the active regulatory interactions contributing to each correlation component, as explained in Eqs.~(\ref{eq4}-\ref{eq6}).
(a) Correlation between X and Y ($\eta_{xy}^2$) is influenced by the regulatory path X$\multimap$Y. (b) Correlation between X and Z ($\eta_{xz}^2$) arises due to direct (X$\multimap$Z) and indirect (X$\multimap$Y$\multimap$Z) pathways. (c) Correlation between Y and Z ($\eta_{yz}^2$) is achieved through three pathways. First, via the regulatory interaction Y$\multimap$Z; second, through the regulatory interaction X$\multimap$Y$\multimap$Z; and third, via interactions X$\multimap$Y and X$\multimap$Z (cross).
\textit{Decomposed output noise}: Colored dashed arrows trace noise propagation routes and the corresponding colored edges denote the active regulatory paths for each component, as described in Eq.~(\ref{eq7}).
(d) $\eta_{z,d}^2$ is accumulated at Z due to direct path X$\multimap$Z. (e) $\eta_{z,ind}^2$ is accumulated at Z due to indirect path X$\multimap$Y$\multimap$Z. (f) $\eta_{z, cross}^2$ is accumulated at Z due to the collective action (cross-interaction) of both direct and indirect path, i.e., $\rm \{X\multimap Z\} \cup \{X \multimap Y \multimap Z\}$. 
The gray lines in the figure stand for inactive edges.
\label{f2}
}
\end{figure*}

The interaction X$\multimap$Y generates the correlation between X and Y, $\eta^2_{xy}$ (Eq.~(\ref{eq4}) and Fig.~\ref{f2}a) due to the regulatory function $f_{yx}^\prime$. Similarly, the correlation between X and Z ($\eta_{xz}^2$) is achieved through two avenues: one due to the interaction along the direct path X$\multimap$Z ($f_{zx}^\prime$) and the other due to the interaction along the indirect path X$\multimap$Y$\multimap$Z ($f_{yx}^\prime f_{zy}^\prime$), as seen in Eq.~(\ref{eq5}) and Fig.~\ref{f2}b. These two partial correlations are denoted by $\eta_{xz,d}^2$ and $\eta_{xz,ind}^2$, respectively.

The correlation between Y and Z ($\eta_{yz}^2$) is developed due to three partial correlations-- $\eta_{yz,ind_1}^2$, $\eta_{yz,ind_2}^2$, and $\eta_{yz, cross}^2$ (Eq.~(\ref{eq6})). Here, $\eta_{yz,ind_1}^2$ corresponds to the regulation Y$\multimap$Z ($f_{zy}^\prime$), while $\eta_{yz,ind_1}^2$ reflects the indirect path X$\multimap$Y$\multimap$Z ($f_{yx}^{\prime^2} f_{zy}^\prime$), as seen in Eq.~(\ref{eq6}) and Fig.~\ref{f2}c. However, $\eta_{yz, cross}^2$ arises from the interactions X$\multimap$Y and X$\multimap$Z ($f_{yx}^\prime f_{zx}^\prime$). We denote this term as the ``cross-interaction correlation" (Fig.~\ref{f2}c), since it is the result of the partial contributions from both direct and indirect pathways. These three partial correlations provide a complete description of how the correlation between Y and Z develops due to the regulatory pathways. This interpretation of correlation patterns holds irrespective of the structure and logic of FFL.

\subsection*{The cross-interaction noise is governed by partial correlations}

Given the covariance decomposition that shows the correlation pattern in FFLs, we reformulate the noise decomposition equation for Z, Eq.~(\ref{eq3}), in terms of the partial correlations, as presented in Eq.~(S13),
\begin{eqnarray}
    \eta_z^2 &=& 
    \underbrace{\frac{1}
    {\langle z\rangle}
    }_{\eta_{z,int}^2}
    +
    \underbrace{
    \frac{f_{zx}^\prime \langle x\rangle}{\beta_z \langle z\rangle} \eta_{xz,d}^2
    }_{\eta_{z,d}^2:~ \rm X \multimap Z}
    +
    \underbrace{
    \frac{f_{zy}^\prime \langle y\rangle}{\beta_z \langle z\rangle} \left( \eta_{yz,ind_1}^2 + \sigma_{yz,ind_2}^2 \right)
    }_{\eta_{z,ind}^2:~ \rm X \multimap Y \multimap Z}
    \nonumber \\
    && 
    + 
    \underbrace{
    \frac{f_{zx}^\prime \langle x\rangle}{\beta_z \langle z\rangle} \eta_{xz,ind}^2
    +
    \frac{f_{zy}^\prime \langle y\rangle}{\beta_z \langle z\rangle} \eta_{yz,cross}^2
    }_{\eta_{z,cross}^2:~ \rm \{X \multimap Z\} \cup \{X \multimap Y \multimap Z\}}.
    \label{eq7}
\end{eqnarray}

\noindent In the above expression, the noise due to the direct pathway, $\eta_{z,d}^2$, is dictated by $f_{zx}^\prime$ and the partial correlation $ \eta_{xz,d}^2$ (Fig.~\ref{f2}d). The noise caused by the indirect pathway, $\eta_{z,ind}^2$, is dictated by $f_{zy}^\prime$, and the partial correlation $\eta_{yz,ind_1}^2$ and $\eta_{yz,ind_2}^2$ (Fig.~\ref{f2}e). 
Similarly, the fourth term $\eta_{z,cross}^2$, has two components. The first component involves $f_{zx}^\prime$ and the partial correlation $\eta_{xz, ind}^2$. The second component comprises $f_{zy}^\prime$ and the partial correlation $\eta_{yz, cross}^2$ (Fig.~\ref{f2}f). Taking into account the explicit expressions of $\eta_{xz,ind}^2$ and $\eta_{yz, cross}^2$ from Eqs.~(\ref{eq5}) and (\ref{eq6}) and rearranging the cross-interaction noise, it is clear that 
\begin{equation}
    \eta_{z, cross}^2 \propto f_{yx}^\prime f_{zx}^\prime f_{zy}^\prime.
    \label{eq5a}
\end{equation}

\noindent Therefore, in addition to the regulatory functions, the partial correlations are not merely mathematical terms. They are able to identify the precise regulatory origins of noise propagation by delineating which pathways are responsible for specific correlations. To highlight their functional significance, the individual components of $\eta_{z,cross}^2$ are shown as functions of the noise of X in Fig.~S2 (AND regulation) and Fig.~S3 (OR regulation) for C1- and I1-FFLs.

Importantly, the cross-interaction noise $\eta_{z, cross}^2$ serves as a unified descriptor of interactions between the direct and indirect pathways. Its value depends on the presence of all three edges in the FFL. If any regulatory link is removed, the corresponding derivative $f_{\cdots}^\prime=0$, resulting in $\eta_{z, cross}^2=0$ (Eq.~\ref{eq5a}). This reflects the breakdown of the parallel pathway structure, eliminating the source of pathway interaction. Mechanistically, the cross-interaction noise originates from two key correlation channels: the indirect path X$\multimap$Y$\multimap$Z, captured by $\eta_{xz,ind}^2$, and the correlation between Y and Z due to X, $\rm \{X \multimap Y\} \cup \{X \multimap Z\}$, captured by $\eta_{yz, cross}^2$.

\begin{figure*}[!t]
\includegraphics[width=2.0\columnwidth,angle=0]{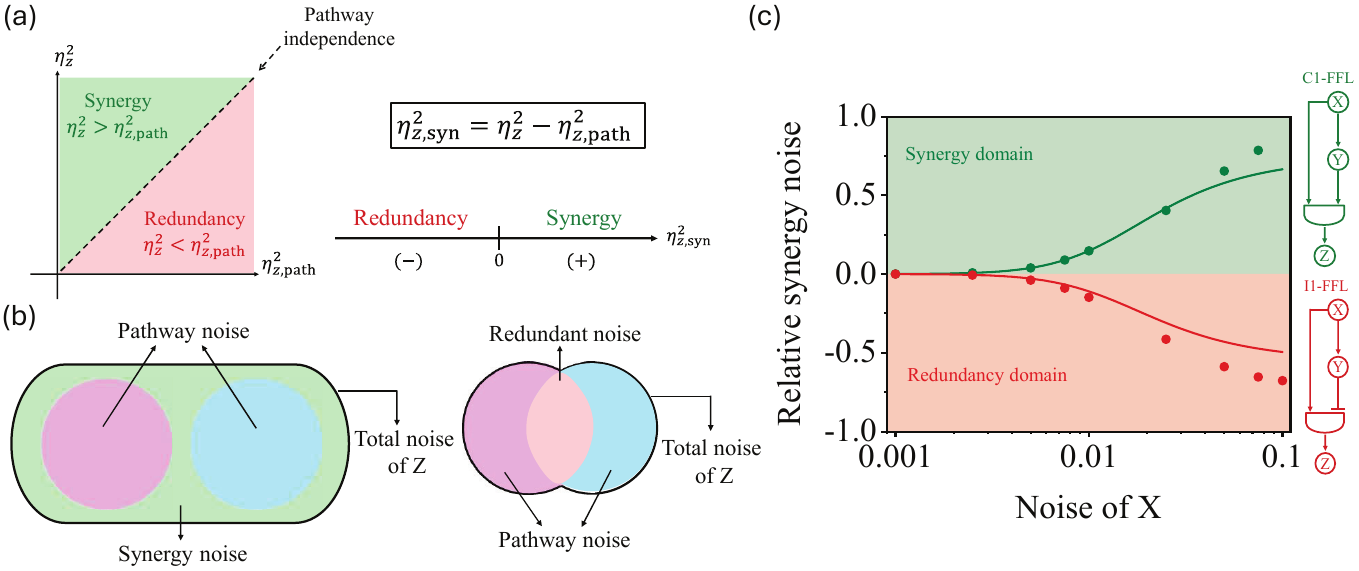}
\caption{\textbf{Characterization of cross-interaction noise as synergy and redundancy.}
(a) Schematic presentation of synergy and redundancy in noise propagation. The difference between total noise and pathway noise is defined as the synergy noise, $\eta_{z, syn}^2$. Although we refer to this as synergy noise, its sign determines the interpretation: a positive value indicates synergy (noise amplification), whereas a negative value indicates redundancy (noise suppression). The diagonal line represents the condition of pathway independence, where total noise equals pathway noise.
(b) Conceptual illustration of synergy and redundancy in noise space, demonstrating how total noise at the output Z is influenced by the interaction between the two pathways. 
(c) Relative synergy noise of C1- and I1-FFLs for AND regulation as a function of the noise of X. C1-FFL exists in the synergy domain, while I1-FFL exists in the redundancy domain. We note that the noise of X is defined as $\eta_x^2=1/\langle x\rangle$, where we set different values of $\langle x\rangle$ to get a range of $\eta_x^2$. Plots are generated using the mathematical framework and parameter sets provided in SI. The lines are due to theoretical calculation, while the points are generated by stochastic simulation algorithm \cite{Gillespie1977}.
\label{f3}
}
\end{figure*}

\subsection*{Cross pathway interaction: Synergy and redundancy}

To better understand the significance of cross-interaction noise in the noise performance of FFLs, we introduce the notion of ``pathway noise"-- defined as the sum of intrinsic, direct, and indirect noise components. We note that the intrinsic noise arises from random birth–death dynamics and is not directly influenced by regulatory structure. However, for consistency in our decomposition, we treat intrinsic noise as a constant scaling factor and include it within the pathway noise definition in the rest of the study. Consequently, we recast the overall noise of Z, represented in Eq.~(\ref{eq3}) and Eq.~(\ref{eq7}), as,
\begin{equation}
    \eta_z^2=\eta_{z,path}^2+\eta_{z,cross}^2,
    \label{eq8}
\end{equation}

\noindent where $\eta_{z,path}^2=\eta_{z,int}^2+\eta_{z,d}^2+\eta_{z,ind}^2$. 

Beyond the pathway noise, the cross-interaction noise signifies a statistical dependence between the two pathways. We interpret it as a measure of the interaction between the two pathways in noise propagation, its sign determining whether the interaction is synergistic or redundant. When $\eta_{z, cross}^2 > 0$ as observed in C1-FFL (see Fig.~S2a for AND regulation and Fig.~S3a for OR regulation), the two pathways collectively amplify fluctuations beyond their individual contributions. In contrast, when $\eta_{z, cross}^2 < 0$, as shown in I1-FFL (see Fig.~S2a for AND regulation and Fig.~S3a for OR regulation), the pathways dampen the fluctuations of each other, leading to a reduced variability of the output and reflecting redundancy. We note that positive $\eta_{z,\text{cross}}^2$ values can also occur in other coherent FFLs with both AND and OR regulation, while negative values can similarly arise in incoherent FFLs across both logic types--highlighting the generality of this behavior.

To clarify this interpretation, we redefine the cross-interaction noise $\eta_{z, cross}^2$ as \textit{synergy noise} and express it as the difference between the total noise and the summed individual pathway contributions (Eq.~(\ref{eq8})), that is, 
\begin{equation}
    \eta_{z,syn}^2 = \underbrace{\eta_z^2}_{\text{Total noise}} - \underbrace{\eta_{z,path}^2}_{\substack{\text{Individual contributions} \\ \text{of the two pathways}}}
    \label{eq9}
\end{equation}

\noindent Although we refer to this difference as \textit{synergy noise}, we emphasize that its sign determines its functional interpretation. When $\eta_{z, syn}^2 = 0$, the total noise is simply equals the individual pathway contributions, reflecting \textit{pathway independence}. However, When $\eta_{z, syn}^2$ is positive, it retains its original interpretation as \textit{synergy}, which means that the pathways propagate more noise than the sum of their individual contributions. However, when $\eta_{z, syn}^2$ is negative, it indicates \textit{redundancy}, where the two pathways propagate less noise than the sum of their individual contributions. We note that a similar notion of synergy (redundancy) appears in neuroscience, where interactions among multiple inputs convey more (less) information than the sum of their parts \cite{Schneidman2003}.  In analogy, our framework interprets such collective effects at the level of noise propagation, offering a systems-level view of how multiple regulatory pathways interact to shape output variability in genetic networks.
This formulation offers a natural and quantitative method for evaluating the degree of pathway dependence in noise propagation, demonstrating how parallel regulatory structures affect the noise characteristics of the output gene Z. These concepts are illustrated in Fig.\ref{f3}a, which shows the correlation between total noise and pathway noise, and in Fig.\ref{f3}b, which visualizes synergy and redundancy in noise space.

We now define \textit{relative synergy noise} as the ratio of synergy noise to the pathway noise, 
\begin{equation}
    {\rm Relative~ synergy~ noise} = \frac{\eta_{z,syn}^2}{\eta_{z,path}^2}.
    \label{eq10}
\end{equation}

\noindent Relative synergy noise provides a normalized measure of pathway interaction strength, enabling direct comparison across different types of FFL. It highlights the extent to which synergistic or redundant interactions dominate over intrinsic and pathway-specific contributions. When both FFL pathways operate independently, the relative synergy noise is zero. A positive value indicates a contribution from synergy, while a negative value suggests redundancy relative to the individual contribution to the pathway. These scenarios-- positive for C1-FFL and negative for I1-FFL are illustrated in Fig.~\ref{f3}c for AND regulation. A similar pattern holds for C1- and I-FFLs with OR regulation, as shown in Fig.~S4. We note that analogous patterns can be observed for other coherent and incoherent FFL types with both type of regulations, indicating that relative synergy noise reflects the underlying network logic regardless of the specific FFL architecture.

\begin{figure*}[!t]
\includegraphics[width=2.0\columnwidth,angle=0]{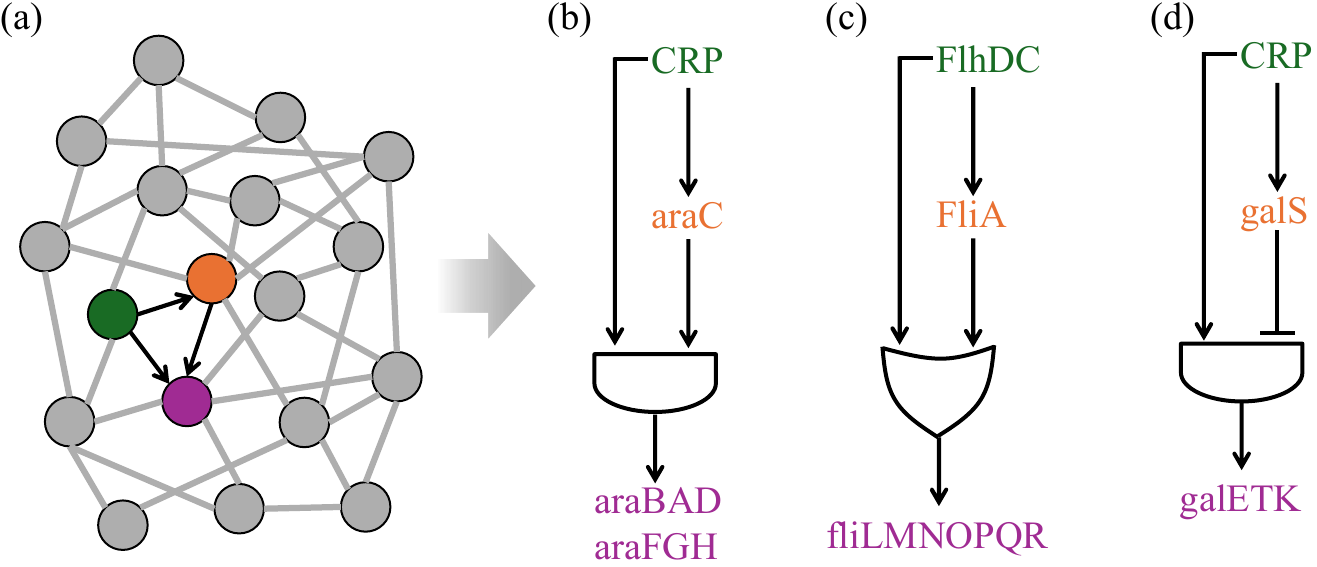}
\caption{\textbf{A representative transcription network of \textit{Escherichia coli} with some examples of FFL systems.} 
(a) The illustrative transcription network. 
(b) An AND-regulated C1-FFL  controlling the arabinose operon, known to implement a sign-sensitive delay \cite{Mangan2003}. 
(c) An OR-regulated C1-FFL in the flagellar gene regulatory system, exhibiting an off-delay response \cite{Kalir2005}. 
(d) An AND-regulated I1-FFL functioning as a response-time accelerator \cite{Mangan2006}.
These examples illustrate how different FFL architectures give rise to distinct dynamical behaviors, which leave identifiable signatures in steady-state noise profiles through synergy and redundancy.
\label{f4}
}
\end{figure*}

\subsection*{Noise synergy and redundancy as fingerprints of regulatory behavior}

We illustrate how synergy and redundancy in noise qualitatively reflect the functional logic of different FFLs. To explore this, we examine several well-characterized examples of FFL systems found in the transcriptional regulatory network of \textit{Escherichia coli} (Fig.~\ref{f4}a).

The AND-regulated system involving C1-FFL, as observed in the arabinose operon of \textit{Escherichia coli}, acts as a sign-sensitive delay filter \cite{Mangan2003}. In this system, CRP (X) activates both the intermediate regulator araC (Y) and the target operons araBAD/araFGH (Z). Additionally, araC also activates Z. The expression of Z occurs only when both X and Y are active, effectively filtering out irrelevant signals and responding solely to sustained inputs (Fig.~\ref{f4}b) \cite{Mangan2003, Alon2006}. This simultaneous activity of the two regulatory pathways creates a statistical imprint in the form of synergistic noise in a steady-state, which reflects their interrelated influence on the output.

In comparison, the flagellar gene network of \textit{Escherichia coli} features a C1-FFL where the FlhDC (X) and FliA (Y) proteins independently activate the fliLMNOPQR operon (Z), implementing an OR regulation (see Fig.~\ref{f4}c) \cite{Kalir2005}. This system produces an output expression that persists even after the input has been removed, resulting in an off-delay response. The sustained expression following the withdrawal of the signal is evident in steady-state as a statistical signature, exhibiting synergy noise. The level of synergy observed here is approximately 30$\%$ lower than that of the AND-regulated system (refer to Fig.~\ref{f3}c and Fig.~S4), which aligns with the looser integration logic, while still being functionally relevant.

On the other hand, an example of a I1-FFL is the CRP–galS–galETK system in \textit{Escherichia coli}. In this system, CRP (X) activates both the target operon galETK (Z) and its repressor galS (Y) (Fig. \ref{f4}d) \cite{Mangan2006}. This architecture allows for rapid activation by X, followed by delayed repression by Y, which acts as a mechanism to accelerate response time. In our analysis, the steady-state fluctuations of Z are primarily influenced by redundancy, reflecting the antagonistic yet converging effects of X and Y. Consequently, this noise redundancy serves as a steady-state representation of the adaptive timing characteristics of I1-FFLs.

These examples demonstrate how noise synergy and redundancy in steady-state conditions can act as indicators of the underlying dynamic regulation. Synergy represents the temporal coordination of converging signals, while redundancy reflects overlapping, antagonistic signals that help stabilize the output. By connecting these characteristics with known FFL behaviors, our framework integrates mathematical structure with regulatory function. This enables motif classification and guides future analyses of larger transcriptional networks.

Beyond encoding dynamical regulation, the concepts of noise synergy and redundancy illustrate functional trade-offs in cellular decision-making. Synergy can enhance the phenotypic variability between isogenic cells, thus supporting bet hedging strategies that improve survival under fluctuating conditions \cite{Chalancon2012}. In contrast, redundancy provides a buffering capacity that suppresses extrinsic or potentially harmful fluctuations, thus improving the fidelity and robustness of gene expression responses \cite{Chalancon2012}. As a result, synergy promotes adaptability, although at the cost of precision, while redundancy ensures control and reliability, which may limit flexibility. This highlights a fundamental trade-off influenced by the FFL topology.

\section*{Discussion}
\label{sec3}

In this work, we identify and quantitatively analyze an often overlooked component of noise in FFLs, the cross-interaction noise arising from cross-pathway interactions. By examining inter-gene correlations, we trace its regulatory origin and interpret it as synergy or redundancy in noise propagation. This distinction captures whether parallel regulatory pathways amplify or suppress fluctuations, depending on the type of FFL. Importantly, this synergy–redundancy framework is broadly applicable to any FFL architecture--whether coherent or incoherent, and regardless of whether the inputs regulate the output via AND or OR logic. In essence, the synergy within C1-FFL intensifies the output noise, whereas the redundancy in I1-FFL diminishes it. Redundancy in I1-FFL is due to repression ($\rm Y \dashv X$), which suppresses the output noise levels.

Our synergy–redundancy framework provides a mechanistic basis for interpreting previously observed patterns of noise amplification and suppression in coherent and incoherent FFLs \cite{Dunlop2008, Li2009, Kittisopikul2010, Osella2011, Grigolon2016, Carignano2018}, thus revealing how network topology influences gene expression variability. By qualitatively relating steady-state synergy and redundancy to known dynamical behaviors, we argue that these noise components provide interpretable signatures of the regulatory logic. Moreover, synergy supports adaptability by increasing expression variability, while redundancy improves stability by filtering noise, but at the cost of reduced flexibility \cite{Chalancon2012}. Such features are crucial for the functional roles of FFLs, which may contribute to the evolutionary selection of FFLs.

One can extend this framework to post-transcriptional circuits such as sRNA-mRNA mediated FFLs, where fast RNA–RNA interactions can induce strong cross-path coupling and nonlinear noise propagation. However, because these systems involve both transcriptional control of the sRNA and post-transcriptional regulation of the mRNA, a separate analysis is needed to understand how pathway-specific and synergistic or redundant noise components contribute to output variability. These findings establish our synergy-redundancy framework as a generalizable tool for understanding how regulatory network architecture shapes noise propagation.

Since feedforward loops (FFLs) are common subcircuits within large gene regulatory networks, our decomposition strategy offers a modular and scalable approach to analyze noise propagation in complex systems. This framework allows for local stochastic analysis to inform global network behavior by isolating cross-pathway contributions within individual motifs. This is particularly relevant in eukaryotic systems, where FFLs are often nested or interconnected. By applying synergy-redundancy decomposition in these contexts, we can identify specific types of FFLs that play critical roles in controlling variability, enhancing robustness, or facilitating decision-making in the presence of noise.

\section*{Materials and methods}
\label{sec4}

In a generic feed-forward loop (FFL), the mean-field dynamics of the three gene products, X, Y, and Z, are governed by the ordinary differential equations,
\begin{equation}
    \frac{d \mathbf{n}}{dt} = \mathbf{f} - \mathbf{g},
    \label{eqn11}
\end{equation}

\noindent where $\mathbf{n} = (x,y,z)^T$ denotes the copy number of the respective molecular species. The vector $\mathbf{f} = (f_x,f_y,f_z)^T$ and $\mathbf{g} = (g_x,g_y,g_z)^T$ represent the production and degradation rates of the components, respectively. Eq.~(\ref{eqn11}) is the general form and is applicable to any type of FFL, coherent or incoherent, as well as for any regulation (e.g., AND or OR) of the direct and indirect pathways at the output node. The explicit expressions of $f_i$ and $g_i$ for C1- and I1-FFLs with both AND and OR regulation are reported in Table~S1.

In a fluctuating environment, the dynamics of the FFL can be described by a chemical master equation that governs the time evolution of the joint probability distribution $P(\mathbf{n};t)$,
\begin{eqnarray}
    \frac{dP(\mathbf{n};t)}{dt} &=& \sum_{i = \mathbf{n}} 
    \left[ 
    \left( \mathbb{E}_i^{+1}-1 \right) g_i P(\mathbf{n};t)
    \right. \nonumber \\
    && \left.    +
    \left( \mathbb{E}_i^{-1}-1 \right) f_i P(\mathbf{n};t)
    \right],
    \label{eqn12}
\end{eqnarray}

\noindent where $\mathbb{E}_i^{\pm 1}$ refers to the step operator that when operating increases ($\mathbb{E}_i^{+1}$) or decreases ($\mathbb{E}_i^{-1}$) the number of molecules by 1. The master equation~(\ref{eqn12}) is applicable to all types of FFL, regardless of the regulatory logic (e.g., AND or OR).

Eq.~(\ref{eqn12}) can be solved using van Kampen’s system size expansion (or $\Omega~$expansion) \cite{Elf2003, Hayot2004, Kampen2007}, which yields closed-form expressions of the second-order moments of the steady-state distribution $P(\mathbf{n})$. This procedure systematically approximates the master equation and yields a Lyapunov equation governing the covariance matrix \cite{Nandi2024}. The Lyapunov equation for the FFL system is given by
\begin{equation}
    \mathbf{J}.\mathbf{\sigma} + \mathbf{\sigma}.\mathbf{J}^T + \mathbf{D} = 0,
    \label{eqn13}
\end{equation}

\noindent where, $T$ refers to the matrix transposition, $\mathbf{J}$ is the Jacobian matrix evaluated at steady-state, and $\mathbf{\sigma}$ is the steady-state covariance matrix. The elements of Jacobian are defined as $J_{ij}=\left( \partial/\partial\langle j\rangle \right) \left( \bar{f}_i - \bar{g}_i \right)$, with bars indicating the evaluation in steady-state (Table~S2). The diffusion matrix $\mathbf{D}$ accounts for the Gaussian white noise statistics, with each element as $D_{ij} = \left( \bar{f_i} + \bar{g_i} \right)\delta_{ij}\delta(t-t^\prime) = 2 \bar{g_i} \delta_{ij}\delta(t-t^\prime)$. This definition reflects the crucial assumption of Gaussian white noise that the noise processes for each species are statistically uncorrelated in time and across components and have zero mean \cite{Elf2003, Paulsson2004, Swain2004, Tanase2006, Kampen2007, deRonde2010}. We note that $\bar{f_i} + \bar{g_i} = 2 \bar{g_i}$ can be understood by taking Eq.~(\ref{eqn11}) at steady-state. Since the degradation processes of each molecular component are independent of the types of FFL, we set $g_i = \beta_i i$ for the time-dependent evolution and $\bar{g}_i = \beta_i \langle i\rangle$ for the steady-state condition, where $\beta_i$ refers to the degradation rate constant.

 
\begin{acknowledgments}
MN thanks SERB, India, for the National Post-Doctoral Fellowship [PDF/2022/001807]. 
\end{acknowledgments}



%

\vspace{-0.2cm}


\section*{Supporting Information}

\setcounter{equation}{0}
\setcounter{figure}{0}
\setcounter{table}{0}
\setcounter{page}{9}
\makeatletter

\renewcommand{\thesection}{S\arabic{section}} 
\renewcommand{\thetable}{S\arabic{table}}  
\renewcommand{\thefigure}{S\arabic{figure}} 
\renewcommand{\theequation}{S\arabic{equation}}

Solving the Lyapunov equation~(12) for the covariance matrix, we obtain the following expression of variances and covariances,
\begin{eqnarray}
    \sigma_x^2 &=& \langle x\rangle,
    \label{S4} \\
    \sigma_{xy}^2 &=& \frac{f_{yx}^\prime}{\beta_x+\beta_y} \sigma_x^2,
    \label{S5} \\
    \sigma_{xz}^2 &=& \frac{f_{zx}^\prime}{\beta_x+\beta_z} \sigma_x^2 + \frac{f_{zy}^\prime}{\beta_x+\beta_z} \sigma_{xy}^2,
    \label{S6} \\
    \sigma_y^2 &=& \langle y\rangle + \frac{f_{yx}^\prime}{\beta_y} \sigma_{xy}^2,
    \label{S7} \\
    \sigma_{yz}^2 &=& \frac{f_{zy}^\prime}{\beta_y+\beta_z} \sigma_y^2 + \frac{f_{zx}^\prime}{\beta_y+\beta_z} \sigma_{xy}^2 + \frac{f_{yx}^\prime}{\beta_y+\beta_z} \sigma_{xz}^2,
    \label{S8} \\
    \sigma_z^2 &=& \langle z\rangle + \frac{f_{zx}^\prime}{\beta_z} \sigma_{xz}^2 + \frac{f_{zy}^\prime}{\beta_z} \sigma_{yz}^2,
    \label{S9}
\end{eqnarray}

\noindent where, $\sigma_i^2$ and $\sigma_{ij}^2$ stand for the variance of the $i$-th component and covariance between $i$ and $j$, where $i,j~(\neq i) \in \{ x,y,z \}$. In the above expressions, we use $f_{ij}^\prime = d\bar{f}_i(\cdots)/d\langle j\rangle$.

We define the normalized covariance as $\eta_{ij}^2=\sigma_{ij}^2/[\langle i\rangle \langle j\rangle]$. Substituting this definition into Eqs.~(\ref{S5}),~(\ref{S6})~and~(\ref{S8}) and rearranging yields,
\begin{eqnarray}
    \eta_{xy}^2 &=& \frac{\Gamma_{xy}}{\langle y\rangle} \times f_{yx}^\prime
    \label{eqs10} \\
    \eta_{xz}^2 &=& 
    \underbrace{\frac{\Gamma_{xz}}{\langle z\rangle} \times f_{zx}^\prime}_{\eta_{xz,d}^2}
    +
    \underbrace{\frac{\Gamma_{xyz}^{(1)}}{\langle z\rangle} \times f_{yx}^\prime f_{zy}^\prime}_{\eta_{xz,ind}^2},
    \label{eqs11} \\
    \eta_{yz}^2 &=& 
    \underbrace{\frac{\Gamma_{yz}}{\langle z\rangle} \times f_{zy}^\prime}_{\eta_{yz,ind_1}^2} 
    +
    \underbrace{\frac{\Gamma_{xyz}^{(2)} \langle x\rangle}{\langle y\rangle \langle z\rangle} \times f_{yx}^{\prime^2} f_{zy}^\prime}_{\eta_{yz,ind_2}^2}
    \nonumber \\
    && 
    + \underbrace{\frac{\Gamma_{xyz}^{(3)} \langle x\rangle}{\langle y\rangle \langle z\rangle} \times f_{yx}^\prime f_{zx}^\prime}_{\eta_{yz,cross}^2},
    \label{eqs12}
\end{eqnarray}

\noindent where the under-braced terms represent the partial covariances contributing to the total normalized covariance. Here, $\Gamma_{xy}=1/(\beta_x+\beta_y)$, $\Gamma_{xz}=1/(\beta_x+\beta_z)$, $\Gamma_{yz}=1/(\beta_y+\beta_z)$, $\Gamma_{xyz}^{(1)}=1/[(\beta_x+\beta_y)(\beta_x+\beta_z)]$, $\Gamma_{xyz}^{(2)}=(\beta_x+\beta_y+\beta_z)/[\beta_y(\beta_x+\beta_y)(\beta_x+\beta_z)(\beta_y+\beta_z)]$, and $\Gamma_{xyz}^{(3)}=(2\beta_x+\beta_y+\beta_z)/[(\beta_x+\beta_y)(\beta_x+\beta_z)(\beta_y+\beta_z)]$.

We now define the output noise by $\eta_z^2=\sigma_z^2/\langle z\rangle^2$. Using this definition in Eq.~(\ref{S9}), we write the expression of noise of Z in terms of partial covariances given in Eqs.~(\ref{eqs10}-\ref{eqs12}) as
\begin{eqnarray}
    \eta_z^2 &=& 
    \underbrace{\frac{1}
    {\langle z\rangle}
    }_{\eta_{z,int}^2}
    +
    \underbrace{
    \frac{f_{zx}^\prime \langle x\rangle}{\beta_z \langle z\rangle} \eta_{xz,d}^2
    }_{\eta_{z,d}^2}
    +
    \underbrace{
    \frac{f_{zy}^\prime \langle y\rangle}{\beta_z \langle z\rangle} \left( \eta_{yz,ind_1}^2 + \sigma_{yz,ind_2}^2 \right)
    }_{\eta_{z,ind}^2}
    \nonumber \\
    && + 
    \underbrace{
    \frac{f_{zx}^\prime \langle x\rangle}{\beta_z \langle z\rangle} \eta_{xz,ind}^2
    +
    \frac{f_{zy}^\prime \langle y\rangle}{\beta_z \langle z\rangle} \eta_{yz,cross}^2
    }_{\eta_{z,cross}^2},
    \label{eqs13}
\end{eqnarray}

\noindent where the under-braced terms represent distinct noise sources. The expressions in Eq.~(\ref{eqs13}) are general and hold for any FFL type. However, to apply them specifically to C1- or I1-FFL with particular pathway regulation mechanisms to output-- such as AND or OR-- one must use the corresponding forms of the production vector $\bm{f}$ (see Table~S1).

We note that the noise of X, defined as $\eta_x^2=1/\langle x\rangle$, is used as the tuning parameter in our analysis. We set different values of $\langle x\rangle$ to obtain the range of the tuning parameter upon which we measure different quantities.


\begin{table}[!h]
    \centering
    \caption{\textbf{The elements of $\mathbf{f}$ for different types of FFL}.}
    \begin{tabular}{lll}
       \textbf{Functions}  & \textbf{C1-FFL} & \textbf{I-FFL} \\
       \hline
        $f_x$ (AND and OR) & $\alpha_x$ & $\alpha_x$ \\
        $f_y$ (AND and OR) & $\alpha_y \frac{x}{K_{xy}+x}$ & $\alpha_y \frac{x}{K_{xy}+x}$ \\
        $f_z$ (AND) & $\alpha_z \frac{x}{K_{xz}+x}\frac{y}{K_{yz}+y}$ & $\alpha_z \frac{x}{K_{xz}+x}\frac{K_{yz}}{K_{yz}+y}$ \\
        $f_z$ (OR) &$\alpha_z \left( \frac{x}{K_{xz}+x}+\frac{y}{K_{yz}+y} \right)$ &$\alpha_z \left( \frac{x}{K_{xz}+x}+\frac{K_{yz}}{K_{yz}+y} \right)$ \\
        \hline
        $\bar{f}_x$ (AND and OR) & $\alpha_x$ & $\alpha_x$ \\
        $\bar{f}_y$ (AND and OR) & $\alpha_y \frac{\langle x\rangle}{K_{xy}+\langle x\rangle}$ & $\alpha_y \frac{\langle x\rangle}{K_{xy}+\langle x\rangle}$ \\
        $\bar{f}_z$ (AND) & $\alpha_z \frac{\langle x\rangle}{K_{xz}+\langle x\rangle}\frac{\langle y\rangle}{K_{yz}+\langle y\rangle}$ & $\alpha_z \frac{\langle x\rangle}{K_{xz}+\langle x\rangle}\frac{K_{yz}}{K_{yz}+\langle y\rangle}$ \\
        $\bar{f}_z$ (OR) & $\alpha_z \left( \frac{\langle x\rangle}{K_{xz}+\langle x\rangle}+\frac{\langle y\rangle}{K_{yz}+\langle y\rangle} \right)$ & $\alpha_z \left( \frac{\langle x\rangle}{K_{xz}+\langle x\rangle}+\frac{K_{yz}}{K_{yz}+\langle y\rangle} \right)$ \\
        \hline
    \end{tabular}
    \label{t1}
\end{table}


\begin{table}[!t]
    \centering
    \caption{\textbf{The kinetic parameters}. Given these values, the parameter $\alpha_i$ is evaluated by considering Eq.~(10) at steady-state.}
    \begin{tabular}{ll}
       Parameter  & Value \\
       \hline
       $\beta_x$ & 0.1 min$^{-1}$ \\
       $\beta_y$ & 1.0 min$^{-1}$ \\
       $\beta_z$ & 10.0 min$^{-1}$ \\
       $K_{xy}$ & 100 molecules/V \\
       $K_{xz}$ & 100 molecules/V \\
       $K_{yz}$ & 100 molecules/V \\  
       \hline
    \end{tabular}
    \label{st2}
\end{table}



\begin{figure*}[!t]
\includegraphics[width=1.5\columnwidth,angle=0]{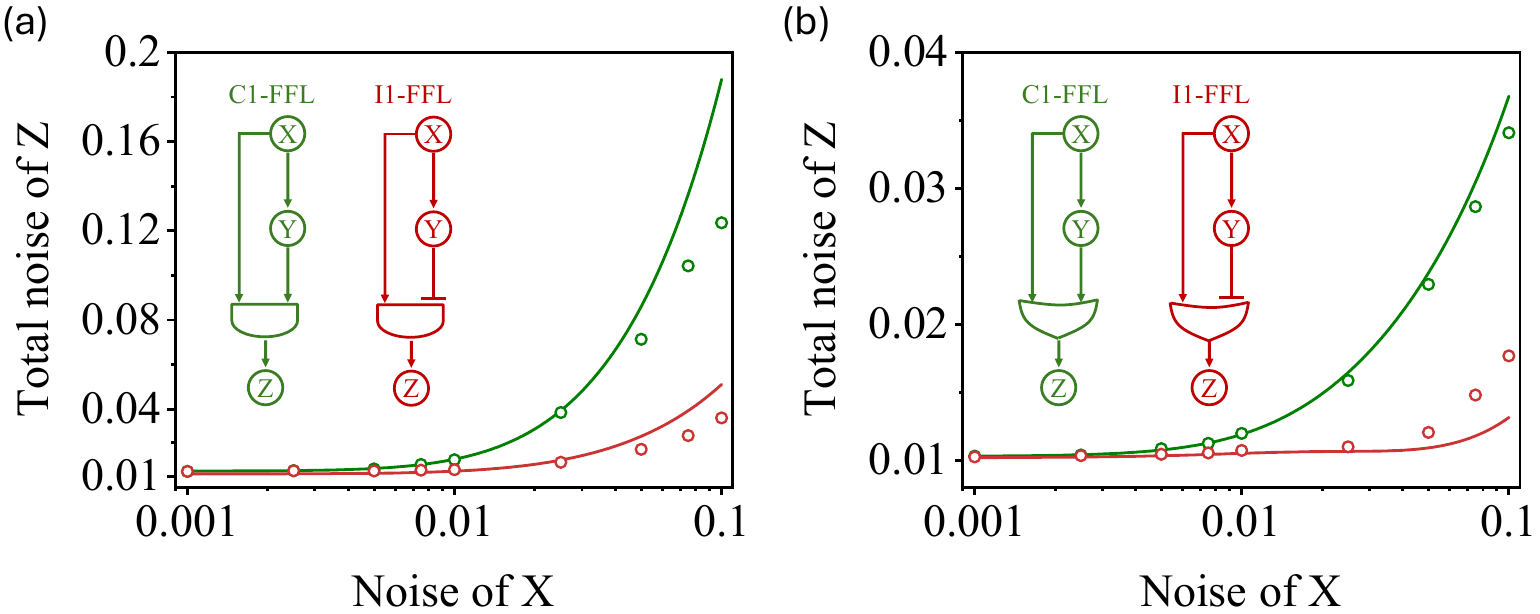}
\caption{\textbf{Behavior of total noise of Z for different FFL structures}. (a) Total noise of Z for C1-FFL and I1-FFL with AND logic integration between direct and indirect pathways (b) Total noise of Z for C1-FFL and I1-FFL with OR logic integration between direct and indirect pathways. The numerical values of the parameters used to generate this graph are reported in Table~S2. The lines are due to theoretical calculations, while the stochastic simulation algorithm \cite{Gillespie1977} generates the points.
}
\label{sf1}
\end{figure*}

\begin{figure*}[!t]
\includegraphics[width=1.2\columnwidth,angle=0]{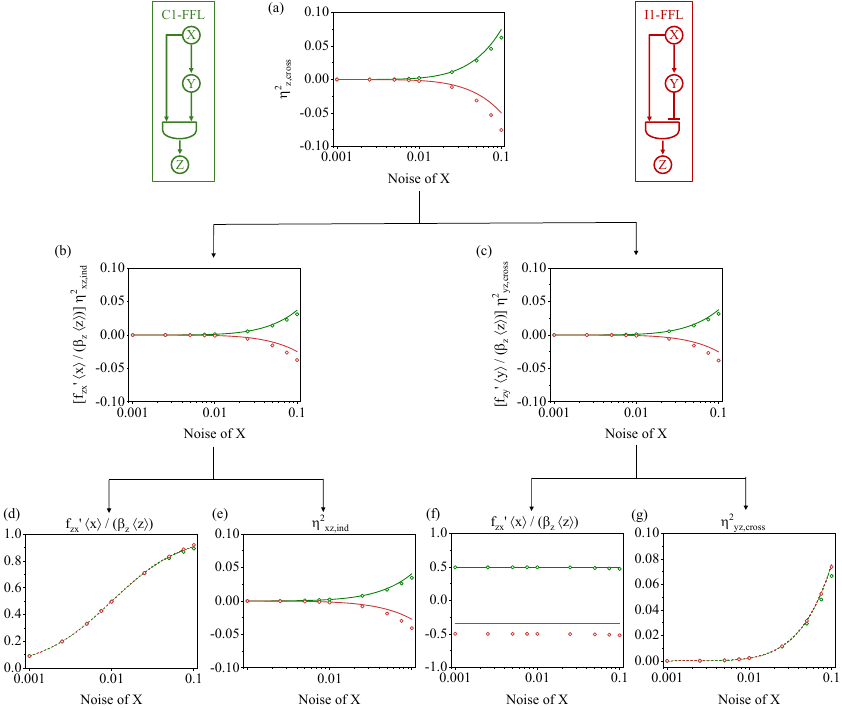}
\caption{\textbf{Variation of additional noise and its constituent parts for C1- and I1-FFLs with AND logic integration as a function of noise of X}. (a) $\eta_{z,cross}^2$. (b-c) The two components of $\eta_{z,cross}^2$. (d-g) Individual parts of each component. We use a similar parameter space to generate these plots as described in Fig.~\ref{f1}. The lines are due to theoretical calculations, while a stochastic simulation algorithm \cite{Gillespie1977} generates the points.
}
\label{sf2}
\end{figure*}

\begin{figure*}[!t]
\includegraphics[width=1.2\columnwidth,angle=0]{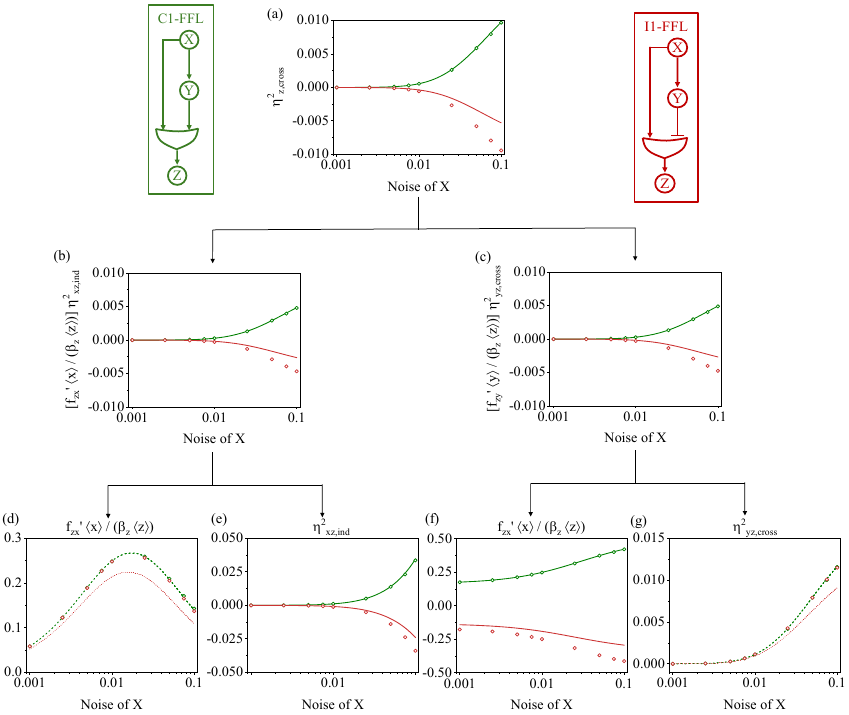}
\caption{\textbf{Variation of additional noise and its constituent parts for C1- and I1-FFLs with OR logic integration as a function of noise of X}. (a) $\eta_{z,cross}^2$. (b-c) The two components of $\eta_{z,cross}^2$. (d-g) Individual parts of each component. We use a similar parameter space to generate these plots as described in Fig.~\ref{f1}. The lines are due to theoretical calculations, while a stochastic simulation algorithm \cite{Gillespie1977} generates the points.
}
\label{sf3}
\end{figure*}

\begin{figure*}[!t]
\includegraphics[width=1.2\columnwidth,angle=0]{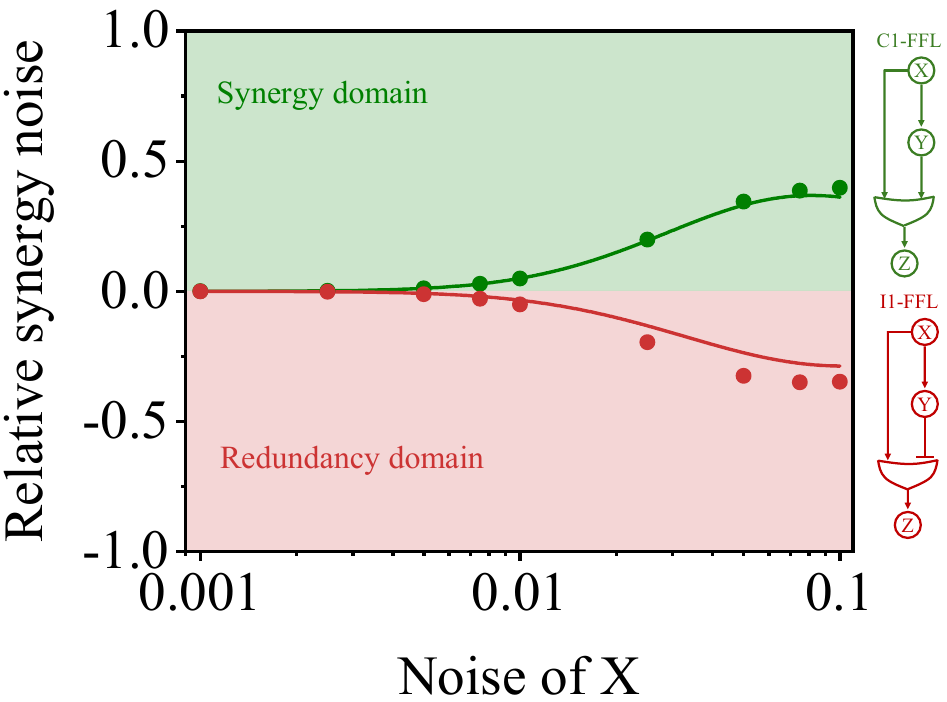}
\caption{\textbf{Relative synergy noise for C1- and I1-FFLs with OR logic integration as a function of the noise of X}. The lines are due to theoretical calculations, while a stochastic simulation algorithm \cite{Gillespie1977} generates the points.
}
\label{sf4}
\end{figure*}

\end{document}